\documentclass[journal]{IEEEtran}
\ifCLASSINFOpdf

\else

\fi

\usepackage{graphicx,amsmath,amssymb,cite}
\usepackage{color}
\usepackage{amsfonts}
\usepackage{amsmath}
\usepackage{graphicx}
\usepackage{amssymb}
\usepackage{stfloats}
\usepackage{mathrsfs}
\usepackage{subfloat}  % new
\usepackage{subfig}  % new

\begin{document}
	%\title{Directional Modulation, DOA Measurement,  and Precise Secure Wireless transmission for 5G and Beyond Using Massive MIMO}
	\title{{RIS Deployment in 6G: Do Low-Rank or Full-Rank Channels Benefit More from DoF Enhancement?}}
	\author{Yongqiang~Li,~Feng Shu,~Maolin~Li,~Qingqing~Wu,~Ke~Yang,~Bin~Deng,~Xuehui~Wang,\\~Fuhui~Zhou, and~Cunhua~Pan 
		\thanks{This work was supported in part by the National Natural Science Foundation of China under Grant U22A2002, and  by the Hainan Province Science and Technology Special Fund under Grant ZDYF2024GXJS292; in part by the Scientific Research Fund Project of Hainan University under Grant KYQD(ZR)-21008; in part by the Collaborative Innovation Center of Information Technology, Hainan University, under Grant XTCX2022XXC07; in part by the National Key Research and Development Program of China under Grant 2023YFF0612900. (Corresponding author: Feng Shu.) }
			\thanks{Yongqiang~Li, Maolin Li, Ke~Yang, and~Bin~Deng are with the School
			of Information and Communication Engineering, Hainan University, Haikou, 570228, China (e-mail: 303513161@qq.com, limaolin0302@163.com, 23220854100060@hainanu.edu.cn, d2696638525@126.com).}
			\thanks{Feng Shu is with the School of Information and Communication Engineering, Hainan University, Haikou 570228, China, and also with the School of Electronic and Optical Engineering, Nanjing University of Science and Technology, Nanjing 210094, China (e-mail: shufeng0101@163.com).}
			\thanks{Qinging Wu is with the Department of Electronic Engineering, Shanghai Jiao Tong University, Shanghai 200240, China (e-mail: qingqingwu@sjtu.edu.cn).}
			\thanks{Xuehui~Wang is with the School of Mathematics and Statistics, Hainan Normal University, Haikou 571158, China (e-mail: wangxuehui0503@163.com).}
			\thanks{ Fuhui Zhou is with the College of Artificial Intelligence, Nanjing University of Aeronautics and Astronautics, Nanjing 210016, China (e-mail: zhoufuhui@ieee.org).}
			\thanks{Cunhua Pan is with the National Mobile Communications Research Laboratory, Southeast University, China (e-mail: cpan@seu.edu.cn).}
}

\maketitle

\begin{abstract}
	Reconfigurable intelligent surface (RIS), as an efficient tool to improve receive signal-to-noise ratio,  extend coverage and create more spatial diversity, is viewed as a most promising technique for the future {Internet of Things (IoT) networks}. RIS is very suitable for a special wireless scenario with wireless link between BS and {IoT devices} being completely blocked, i.e., no link. In this paper, we extend its applications to a general scenario, i.e., rank-deficient channel, particularly some extremely low-rank ones such as no link, and line-of-sight (LoS, rank-one). {Several important IoT application scenarios inherently exhibit low-rank channel characteristics, including low-altitude, satellite, UAV, marine, and deep-space communications.} In such a situation, it is found that RIS may make a dramatic degrees of freedom (DoF) enhancement over no RIS. {By exploiting a distributed RIS deployment strategy, the channel from the BS to the IoT device in a LoS scenario can be transformed from a low-rank state (e.g., rank zero or one) to full-rank, thereby substantially enhancing the available DoF.} This will achieve an extremely rate improvement via spatial parallel multiple-stream transmission from BS to IoT device. In this paper, we present a complete review of making an in-depth discussion on DoF effect of RIS.

\end{abstract}

\section{RIS-aided Wireless Network: concept and system}

{The advancement of the Internet of Things (IoT) is enabling the vision of a massively connected intelligent world. While this brings unprecedented connectivity, it also introduces critical challenges such as high power consumption and security vulnerabilities. Moreover, the number of devices that can be supported is often constrained by limited channel quality~\cite{Liu2023}, for instance, due to blockages or low-rank conditions. To address these issues, reconfigurable intelligent surfaces (RIS), with their inherent capability for spatial signal manipulation, have emerged as a key enabling technology.}

{Power consumption reduction and security enhancement have been extensively studied in the existing literature~\cite{Wu2019}. For instance, a random subcarrier scheme based on directional modulation (DM) was proposed in~\cite{Shu2017Secure}, enabling secure and precise communication with multiple IoT devices. Enhancing the multi-stream transmission capability of IoT networks is pivotal for supporting massive device access.}
In~\cite{Shu2021}, a RIS-assisted DM network was proposed to transmit two confidential data streams from base station to the IoT device, achieving twice the secrecy rate of single-stream transmission in the high signal-to-noise ratio (SNR) region.
In~\cite{Yang2024}, a distributed RISs-assisted transmission rate enhancement method in LoS channel was studied to increase channel rank. By dividing a large RIS into multiple small RISs for multi-stream transmission, the transmission rate was improved compared to a large RIS-assisted transmission. {Additionally, RIS deployment at cell edges alleviates inter-cell interference \cite{Pan2020}, while its integration with relay stations expands network coverage and increases transmission rates \cite{Wang2022}, thereby enhancing IoT device access capability.}
\begin{figure}[b]
	\centering
	\includegraphics[width=0.40\textwidth, trim = 20 20 10 30,clip]{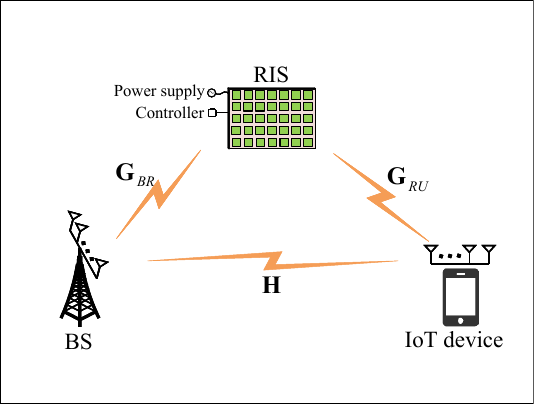}\\
	\caption{Diagram of RIS-aided Wireless Network.}\label{Fig-1}
\end{figure}

{In addition to the aforementioned advantages, the deployment of RIS in IoT networks has been proven to enhance the degrees of freedom (DoF), with analyses conducted for diverse scenarios.} A joint small-scale spatiotemporal correlation model was proposed and analyzed for DoF under isotropic scattering in~\cite{Sun2021}. {It has been demonstrated in \cite{Zheng2023, Liu2024} that the DoF gain is related to the number of antennas and the achievable rate.} Based on information theory, the DoF of the joint and independent transmission of information for RIS and transmitter was investigated in\cite{Cheng2024}. For the rank-deficient channel, when the rank of the channel from the transmitter to the receiver is large and/or the rank of the channel from the transmitter to the RIS is small, the DoF can be significantly improved~\cite{Chae2023}. Reference~\cite{Wang2024} studied the reconfigurable distributed antenna and reflecting surface assisted MIMO system and obtained additional DoF through mode selection of each reflection element. To fully exploit the DoF of RIS, a multi-layer refractive RIS-assisted receiver was proposed~\cite{An2024}, and a robust optimization framework of a low-complexity synchronous wireless information and power transmission scheme was designed to enhance receiver performance. {To clearly demonstrate the DoF improvement enabled by RIS deployment in IoT networks, this work investigates the DoF and achievable rate of two representative low-rank channels (e.g., blocked and LoS links). With RIS assistance, the rank of these channels can be elevated from zero or one to full, thereby achieving significant rate gains.}
{The key contributions are summarized as follows:
\begin{itemize}
	\item The enhancement of the DoF by RISs in IoT networks over low-rank channels is investigated in this paper. By deploying RISs, the low-rank channel can be elevated to full rank, thereby enabling multi-stream parallel transmission and significantly improving the capacity of IoT systems.
	\item To enhance the communication capability of IoT devices in low-rank channels, this work investigates cooperative beamforming using distributed RISs, applicable to low-altitude, satellite, UAV, marine, and deep-space communications. Simulation results verify that the proposed scheme can effectively enhance the DoF, thereby achieving significantly higher achievable rates.
\end{itemize}
}
\section{System Model and DoF Discussion}
Fig.~\ref{Fig-1} shows a typical three-node wireless network aided by RIS. In this figure, the number of antennas at BS, the number of RIS elements, and the number of antennas at {IoT device} are $M$, $N$ and $K$, respectively. The corresponding channel matrices from BS to RIS, RIS to IoT device, and BS to IoT device are denoted as $\mathbf{G}_{BR}$, $\mathbf{G}_{RU}$ and $\mathbf{H}$, respectively. Meanwhile,	the channel phase shifting matrix is defined as $\mathbf{\Theta}$.
\begin{figure*}[t]
	\centering
	%\begin{subfigures}
		\subfloat[DoF = 1.]{\label{fig:1(f)}
			\includegraphics[width=2in, trim = 20 40 10 20,clip]{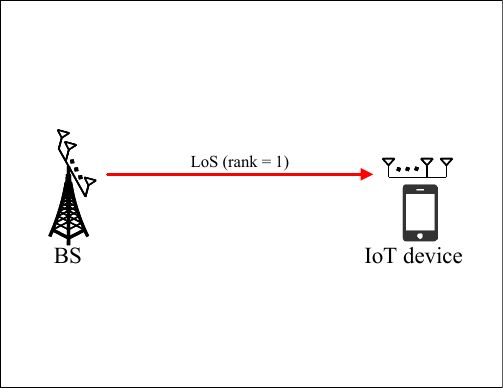}}\hfil
		\subfloat[DoF = 2.]{\label{fig:1(a)}
			\includegraphics[width=2in, trim = 20 10 10 20,clip]{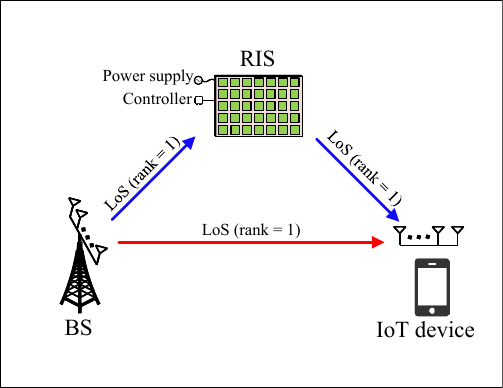}}	\hfil	
		\subfloat[DoF = K.]{\label{fig:1(b)}
			\includegraphics[width=2in, trim = 20 10 10 20,clip]{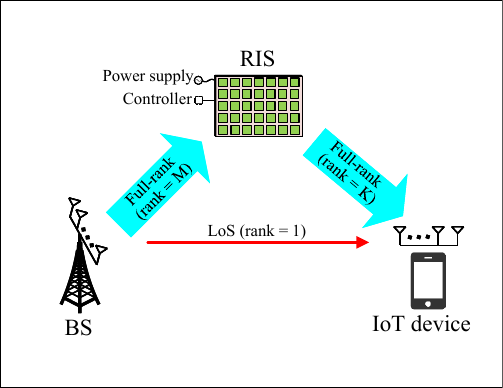}}\\
%		\subfloat[DoF = 0.]{\label{fig:1(c)}
%			\includegraphics[width=2in, trim = 20 50 10 20,clip]{Figs/fig2d.pdf}}\hfil
%		\subfloat[DoF = 1.]{\label{fig:1(d)}
%			\includegraphics[width=2in, trim = 20 10 10 20,clip]{Figs/fig2e.pdf}}\hfil
%		\subfloat[DoF = K.]{\label{fig:1(e)}
%			\includegraphics[width=1.8in, trim = 20 10 10 20,clip]{Figs/fig2f.pdf}}\\
\caption{Diagram of deploying RIS to increase system DoF: (a) rank = 1; (b) Increase from rank = 1 to rank = 2; (c)  Increase from rank = 1 to rank = $K$.}\label{fig:2}	%\end{subfigures}
\end{figure*}
%\begin{figure*}[htbp]
%	\centering
%	\begin{minipage}{0.39\linewidth}
%		\centering
%		\includegraphics[width=0.9\linewidth]{fig2_f.eps}
%		\caption{chutian1}
%		\label{chutian1}%文中引用该图片代号
%	\end{minipage}
%	\begin{minipage}{0.39\linewidth}
%		\centering
%		\includegraphics[width=0.9\linewidth]{fig2_f.eps}
%		\caption{chutian2}
%		\label{chutian2}%文中引用该图片代号
%	\end{minipage}
%		\begin{minipage}{0.39\linewidth}
%		\centering
%		\includegraphics[width=0.9\linewidth]{fig2_f.eps}
%		\caption{chutian2}
%		\label{chutian2}%文中引用该图片代号
%	\end{minipage}
%	%\qquad
%	%让图片换行，
%	
%	\begin{minipage}{0.49\linewidth}
%		\centering
%		\includegraphics[width=0.9\linewidth]{fig2_f.eps}
%		\caption{chutian3}
%		\label{chutian3}%文中引用该图片代号
%	\end{minipage}
%	\begin{minipage}{0.49\linewidth}
%		\centering
%		\includegraphics[width=0.9\linewidth]{fig2_f.eps}
%		\caption{chutian4}
%		\label{chutian4}%文中引用该图片代号
%	\end{minipage}
%	\begin{minipage}{0.49\linewidth}
%		\centering
%		\includegraphics[width=0.9\linewidth]{fig2_f.eps}
%		\caption{chutian2}
%		\label{chutian2}%文中引用该图片代号
%	\end{minipage}
%\end{figure*}

 {It is assumed that $M$ and $N$ are much larger than $K$ due to the small size of IoT devices, with $N \ge M$.} In general, the rank of channel matrix $\mathbf{H}$ varies from 0 to $K$.  Similarly, the ranks of channel matrices $\mathbf{G}_{BR}$ and $\mathbf{G}_{RU}$  range from 0 to $M$ and 0 to $K$ in accordance with their wireless environments. {In a richly scattered IoT network, if any two of the three channel matrices are full-rank, then the composite channel matrix $\mathbf{G}_{BR}\mathbf{\Theta}\mathbf{G}_{RU} + \mathbf{H}$ can achieve a full rank of $K$. For example, if $\mathbf{G}_{BR}$ and $\mathbf{G}{RU}$ have elements drawn from an independent and identically distributed (i.i.d.) complex Gaussian distribution with zero mean, then $\text{rank}(\mathbf{G}_{BR}\mathbf{\Theta}\mathbf{G}_{RU} + \mathbf{H}) = K$.}

 {This paper focuses on a prevalent class of rank-deficient channels, exemplified by deep-space, marine, UAV, and satellite communications, where the channel rank is deficient or even extremely low (e.g., close to one). Such channels are characterized by rank$(\mathbf{H})<K$.} Consequently, they require RIS to elevate their rank to the full rank of $K$. In particular, when rank$(\mathbf{H})=0$ or 1, i.e., an extremely low-rank scenario, the enhanced-DoF role of RIS will become a dominant factor to boost the overall system rank in order to achieve a high-rate transmission. The former rank$(\mathbf{H})=0$, i.e. $\mathbf{H}=\mathbf{0}$, means that the channel from BS to IoT device is blocked fully with no direct signal energy received by IoT device. The latter rank$(\mathbf{H})=1$ implies that the channel is LoS channel or additive white Gaussian noise (AWGN) channel. Even when BS and IoT device are employed with multiple antennas, there is only a single message stream transmitted from BS to IoT device in LoS channel. Therefore, the above two channels demand RIS to improve their channel ranks to implement a multiple-stream transmission from BS to IoT device.  {A detailed discussion of achieving this significant rank/DoF enhancement with RIS is provided in the following two sections.}

\section{DoF Improvement and Beamforming in Typical Rank-deficient Channels}
{In this paper, wireless channels with MIMO are divided into two typical channels: rank-deficient and full-rank. In a richly scattered environment, the channel from the BS to the IoT device, as shown in Fig.~\ref{Fig-1}, is considered a full-rank channel.} The conventional LoS channel without reflecting and scattered paths is a rank-deficient channel. 
%An extremely low-rank channel is shown in Fig. 2a. 
In Fig.~2, a far-field RIS-aided wireless network in LoS channel is plotted. 
%The lowest-rank channel is a rank-zero channel, which is completely blocked such that IoT device cannot receive any power from the signal transmitted by BS, called no-link channel in what follows. 
{The channel rank is clearly shown to increase from one (without RIS, Fig. 2a) to two (with RIS, Fig. 2b), on the condition that the BS-RIS and RIS-IoT device channels are LoS and the RIS is not collinear with the BS and the IoT device.}
 If both reflected channel matrices $\mathbf{G}_{BR}$ and $\mathbf{G}_{RU}$ are full-rank with their elements being i.i.d. Gaussian distributions, the channel rank is boosted to $\min(M, N, K)=K$. {In a no-link channel, the introduction of an RIS can progressively increase the channel rank: from 0 to 1, and up to $K$.} In summary, for a rank-deficient channel with rank $k$ more than LoS, its rank is also improved from $k$ to the range from $k+1$ to full-rank ($K$) with the aid of RIS. Additionally, it is particularly pointed out if the channel from BS to IoT device is a typical rich-scattered IoT channel, i.e. full-rank and rank$(\mathbf{H})=K$, then a RIS or even multiple RISs are introduced into such a system, there are no DoF gain achieved in a scenario since rank$(\mathbf{G}_{BI}\mathbf{\Theta}\mathbf{G}_{RU}+\mathbf{H})=K$. However, reflecting array gain and diversity gain created by RIS always exist. 
\begin{figure}[h]
	\centering
	\includegraphics[width=0.40\textwidth, trim = 20 1 20 20,clip]{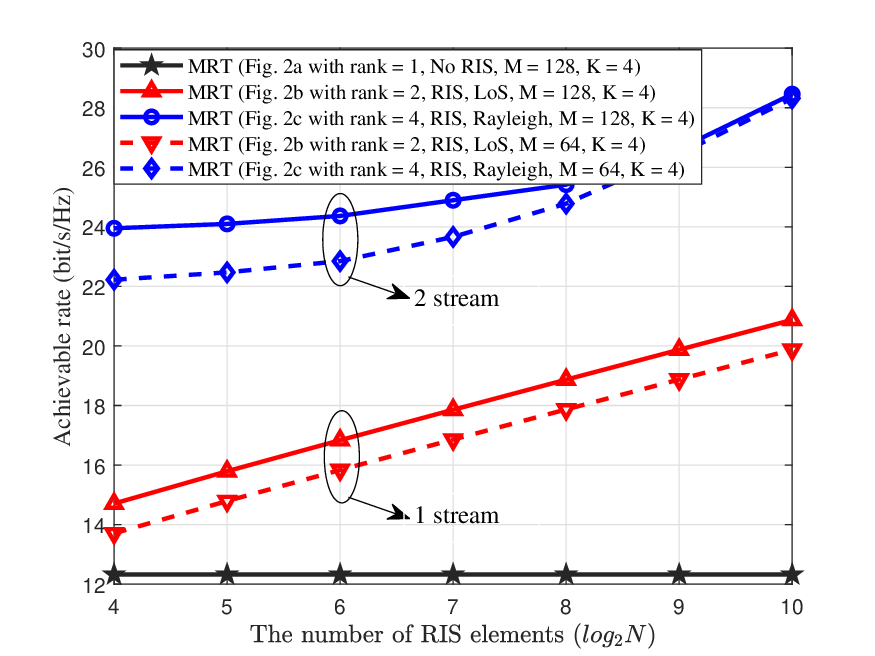}\\
	\caption{Curves of achievable rate versus the number $N$ of RIS elements in LoS channel.}\label{Fig-4}
\end{figure}
\begin{figure}[htp]
	\centering
	\includegraphics[width=0.40\textwidth, trim = 20 1 20 20,clip]{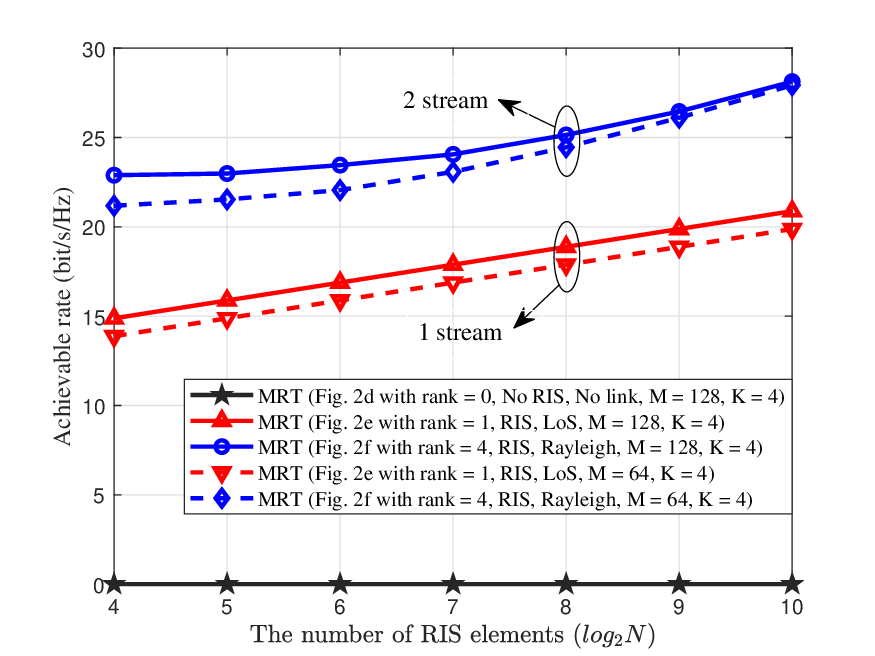}\\
	\caption{Curves of achievable rate versus the number $N$ of RIS elements in scenarios without link.}\label{Fig-5}
\end{figure}

To evaluate the rate performance enhancement achieved by increased DoF due to the introduction of RIS, {numerical simulations are conducted}. System parameters are set as follows:  the distances from BS to RIS, {RIS to IoT device, and BS to IoT device} are 82 m, 28 m, and 100 m, respectively. It is not specifically noted that thermal noise powers at RIS and IoT device are -90 dBm and -70 dBm, respectively. The simulation adopts maximum ratio transmission (MRT) at the transmitter, phase alignment at the RIS, and SNR maximization at the receiver. Fig. \ref{Fig-4} plots the achievable rate in the scenarios with $\text{rank}(\mathbf{H})=1$ both with and without an RIS. Here, the number of RIS elements is 1024 and $K=4$. {When $M=64$, the rates of RIS-aided systems in all-LoS and LoS-plus-Rayleigh fading channels are approximately 1.6 and 2.3 times that of LoS without RIS, primarily due to the increase in system DoF from one to two.} When the number of antennas at BS is increased from 64 to 128, the same trend in rate improvement is observed.

Fig. \ref{Fig-5} demonstrates the curves of achievable rate versus the number of RIS elements  of no link with the aid of RIS, where no link means that the direct channel from BS to {IoT device} is completely blocked. The rate without the RIS is zero due to no link between BS and {IoT device}. By using only single RIS, the maximum DoF of the all-LoS system as shown in Fig. 2e  is one, this means that it may transmit a single bit stream from BS to {IoT device}, which will make a no-link become a single transmission. If both reflected channels $\mathbf{G}_{BR}$ and $\mathbf{G}_{RU}$ are a full-rank MIMO Rayleigh channels as shown in Fig. 2f, the maximum DoF of the system is up to $K$, i.e., full-rank. This implies that and a maximum of $K$ spatial parallel bit streams of data may be transmitted from BS to IoT device, which will harvest a dramatic rate gain.
\section{DoF Boosting and Beamforming via distributed RISs in LoS Channels}
\begin{figure}[htp]
	\centering
	\includegraphics[width=0.40\textwidth, trim = 20 20 10 20,clip]{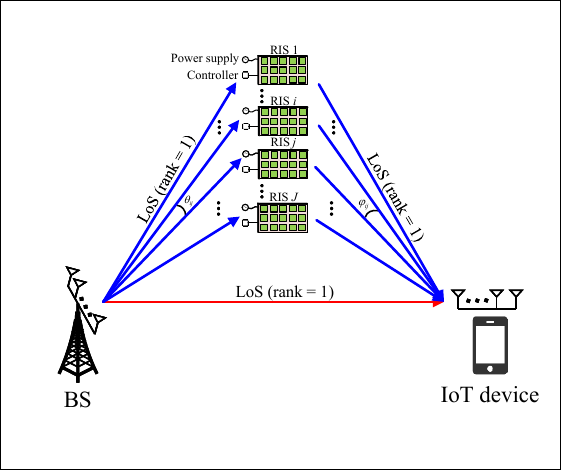}\\
	\caption{Diagram of distributed multi-RIS-aided Wireless Network in LoS channel.}\label{Fig-3}
\end{figure}
In this section, we will discuss a special scenario with both channels from BS to RIS and RIS to {IoT device} as shown in Fig.~\ref{Fig-3} being LoS channels. Is there any way to further boost the DoF value of system? In the following, a distributed multi-RIS concept is proposed to make a dramatic DoF enhancement~\cite{Yang2024} in all-LoS scenario. However, it is particularly noted that the idea is also extended to the cases of all-reflected channels with channel rank being larger than or equal to one. 

In general, a traditional low-rank network has one or two DoFs, which will seriously limit its rate performance. In accordance with basic principle of MIMO system, DoF is vital to boost rate performance. In this section, a novel distributed system model with $J$ RISs  as shown in Fig. \ref{Fig-3} are distributed deployed to create more DoFs. The corresponding channel matrices from BS to RIS $j$ and RIS $j$ to IoT device are denoted as $\mathbf{G}_{BR}^j$ and $\mathbf{G}_{RU}^j$, respectively. If the alignment of spatial features between $\mathbf{H}$ and cascaded channels is low, $J\leq K-1$ RISs with low spatial feature alignment are required. If the spatial features of $\mathbf{H}$ and cascaded channels are aligned, $J\leq K$ RISs are divided. Specifically, the angle $\theta_{ij}$ between $\mathbf{G}_{BR}^i$ and $\mathbf{G}_{BR}^j$ and the angle $\phi_{ij}$ between $\mathbf{G}_{RU}^i$ and $\mathbf{G}_{RU}^j$ should be well designed to satisfy low spatial alignment, and then $\text{rank}(\sum_{j=0}^{J-1}\mathbf{G}_{BR}^j\mathbf{\Theta}^j\mathbf{G}_{RU}^j+\mathbf{H})=K$ is well-conditioned.  The well-conditioned positions of the RISs can be obtained through geometric knowledge and MIMO theory.  According to \cite{Costello2009}, it has been proved  that $\mathbf{G}_{BR}^i\times \mathbf{G}_{BR}^j=\mathbf{0}$  $\mathbf{G}_{RU}^i\times \mathbf{G}_{RU}^j=\mathbf{0}$. Specifically, $\theta_{ij}=\theta_{j}-\theta_{i}\in[0,\pi/2]$ where $\theta_{j}=\arccos(\cos(\theta_{i})+{\lambda l}/{Md})$~, where $l$ is an integer, $\lambda$ denotes the wavelength, and $d$ is the minimum distance between antennas. 
In the extreme situation, the channel matrices  $\mathbf{G}_{BR}^j$ and $\mathbf{G}_{RU}^j$ are rank-one, with each channel having only single eigen-vector. Then, $K=J+1$. Subsequently, a point-to-point $J+1$-stream transmission can be achieved in this rank-deficient channel scenario. Moreover, a significant rate enhancement can be achieved by this model.

%\begin{figure}[htp]
%	\centering
%	\includegraphics[width=0.40\textwidth, trim = 20 1 20 20,clip]{fig_8.eps}\\
%	\caption{The achievable rate versus the number $N_j$ of elements per RIS in scenarios with distributed RISs.}\label{Fig-6}
%\end{figure}
%Fig. \ref{Fig-6} illustrates the curves of achievable rate versus the  number of RIS elements obtained by distributed RISs, where the transmitting beamforming vector, phase adjusting matrix, and receive beamforming vector are designed by null-space-projection, phase alignment, and zero-forcing, respectively. As shown in Fig. \ref{Fig-6}, as the number of distributed RISs grows from one to four, the achieved rate gains are as follows: 1.4, 2.8, 4.2, and 5.6 times over no RIS under the same power-sum constraint. According to this trend, it can be inferred that increasing the number of RISs will further improve the rate performance.
%\begin{figure}[htp]
%	\centering
%	\includegraphics[width=0.40\textwidth, trim = 20 1 20 20,clip]{fig_7a.eps}\\
%	\caption{The achievable rate versus thermal noise power $\sigma_r^2$ at RIS in scenarios with distributed RISs.}\label{Fig-7}
%\end{figure}
\begin{figure}[t]
	\centering
	\subfloat[]{\label{fig:7a}
		\includegraphics[width=3in, trim = 20 1 1 10,clip]{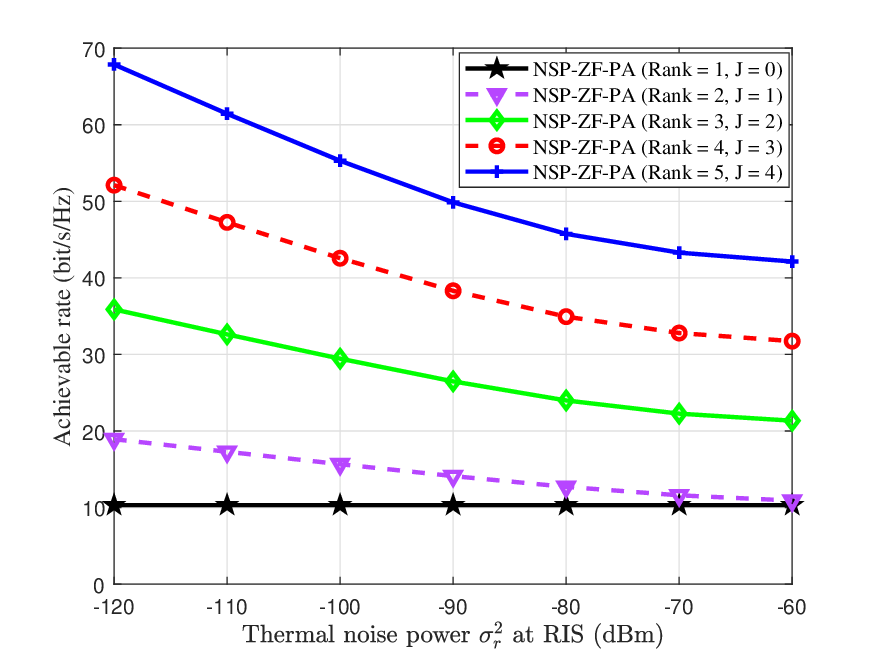}}\hfil
	\subfloat[]{\label{fig:7b}
		\includegraphics[width=3in, trim = 20 1 1 10,clip]{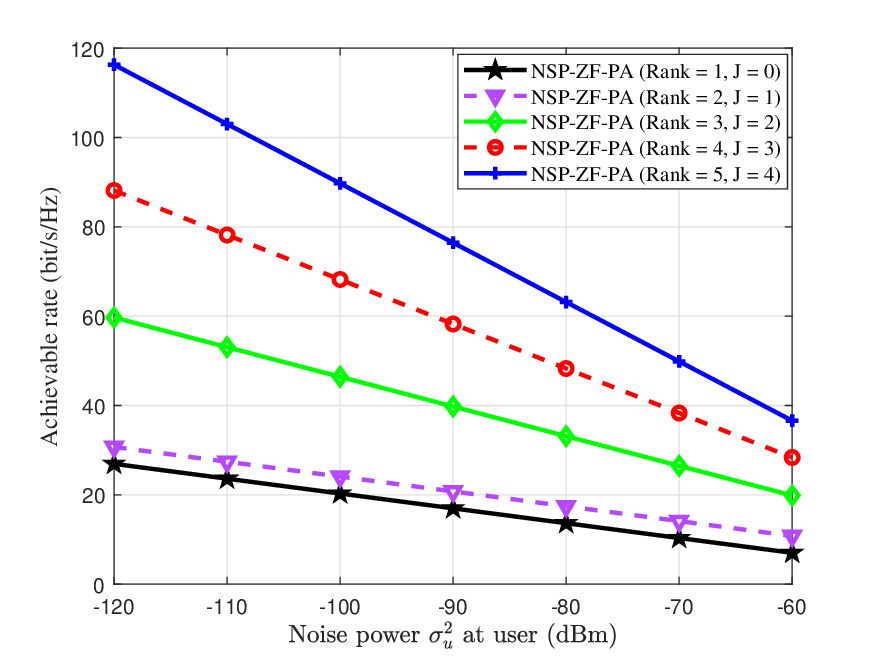}}	
	\caption{(a) The achievable rate versus noise power $\sigma_r^2$ at RIS in scenarios with distributed RISs; (b) The achievable rate versus noise power $\sigma_u^2$ at IoT device in scenarios with distributed RISs.}\label{Fig-7}	
\end{figure}
To evaluate the rate performance enhancement achieved by DoF increasement due to the introduction of distributed RISs, we will present numerical simulations as follows. System parameters are set as follows:  the distances from BS to RIS, RIS to {IoT device}, and BS to IoT device are 82 m, 28 m, and 100 m, respectively; the number of antennas at BS, the number of RIS elements, and the number of antennas at IoT device are $M = 128$, $N = 600$, and $K = 4$, respectively. Fig. \ref{fig:7a} illustrates the curves of achievable rate versus noise power $\sigma_r^2$ at RIS obtained by distributed RISs with $\sigma_u^2 = -70$ dBm, where the transmitting beamforming vector, phase shifting matrix, and receive beamforming vector are designed by null-space-projection, phase alignment, and zero-forcing, respectively. Observing this figure, we find: as noise power $\sigma_r^2$ at RIS increases from -120 dBm to -60 dBm, the achieved rate gains over no RIS gradually decrease and converge to zero. When the number of distributed RISs is greater than or equal to 2, i.e., the system DoF increase by 2 or more, the performance is obviously improved compared with no RIS under high noise power, i.e., $\sigma_r^2=-60$ dBm. As the number of distributed RISs grows from one to four, the achieved rate gains are as follows: about 0.7, 1.9, 2.7, and 3.5 times over no RIS under the same power-sum constraint and $\sigma_r^2=-90$ dBm. According to this trend, it can be inferred that increasing the number of distributed RISs and reducing noise power $\sigma_r^2$ at RIS will further improve the rate performance. 

Fig. \ref{fig:7b} illustrates the curves of achievable rate versus noise power $\sigma_u^2$ at IoT device obtained by distributed RISs with $\sigma_r^2 = -90$ dBm. From this figure, it is seen: as noise power $\sigma_u^2$ at IoT device varies from -120 dBm to -60 dBm, the achieved rate gains over no RIS gradually decrease. 

In summary, we have the following conclusions: (1) Increasing the number of distributed RISs may boost the value of {IoT system} rank or DoF such that a significant rate gain can be attained, (2) low-noise RIS and receiver are also crucial for an active RIS-aided {IoT} network to achieve a high rate gain.

\section{Open Challenging Problems}

%{\color{blue}Maolin is responsible to recommend several open \colorbox{yellow}{problems. For example} multi-user DoF analysis and enhancement, near-field DoF, beamforming and power allocation, multi-user sechluding.}

Due to the above major advantages of RIS and its ability to make a striking DoF improvement in low-rank channels, RIS may extend its application  to several new fields such as
near-sea, near-space, and UAV communications.  However, there still are several open challenging problems to address in RIS fields. Here, we list several important ones of them as follows:
\begin{enumerate}
\item Similar to a movable antenna array, a movable RIS may be introduced to further optimize the space position of RIS over a given 3D area to harvest the performance gains. Additionally, how about its performance upper bounds?
\item In near field, how to compute and evaluate the DoF  or rank value of such a RIS-aided system when the distance from {IoT device} to RIS is less than the Rayleigh distance? It is clear that this problem is hard, however, its DoF upper or lower bounds may be derived as a DoF performance benchmark. Moreover, the corresponding beamforming and phase adjusting methods is also designed to implement a multi-stream near-field transmission from BS to each {IoT device}.

\item In a distributed multi-RIS scenario, how to achieve an optimal matching between RISs and {IoT devices} in order to maximize the sum-rate by using some traditional methods of graph theory in a multiple IoT device situation? Furthermore, similar to scheduling theory, how to make a fair use of RISs per {IoT device}?
\item For a RIS-assisted hybrid beamforming structure, how to jointly design the transmission digital precoding vector, transmission analog weight vector, RIS phase shifting matrix, and reception beamforming vector in order to achieve the maximum DoF of the system? 

\item {How to deploy distributed RISs and fully exploit the available DoF to enhance the quality of service for massive, heterogeneous, and low-capability terminals remains an open challenge.}
\end{enumerate}

\section{Conclusion}

In this article,  the great potential of RIS has been highlighted as a crucial DoF enhancement tool for several typical low-rank future application scenarios such as  marine, low-altitude, UAV, satellite and deep-space communications.  A distributed multi-RIS deployment was proposed to achieve a significant rank enhancement for low-rank wireless situations. There are still several challenging problems ahead to exploit its great potential of the technology. Also, we have raised several new open important research problems. Finally, in our view, a distributed multi-RIS will achieve  wide diverse promising applications in the coming future.
\renewcommand\refname{References}
\bibliographystyle{IEEEtran}
\bibliography{cite}
\ifCLASSOPTIONcaptionsoff
\newpage
\fi
%\bibliographystyle{IEEEtran}
%\bibliography{IEEEfull,cite}
\begin{IEEEbiographynophoto}\\
YONGQING LI is a PhD student in the School of Information and Communication Engineering at Hainan University, Haikou, China. His research interests include massive MIMO, physical layer security, and intelligent reflecting surface.
\end{IEEEbiographynophoto}
\begin{IEEEbiographynophoto}\\
FENG SHU is a professor with  the School of Information and Communication Engineering at Hainan University, Haikou, China, and also with the School of Electronic and Optical Engineering at Nanjing University of Science and Technology, Nanjing, China.  He has been awarded with Mingjian Scholar Chair Professor in Fujian Province, China. He has published more 300 journal  papers on machine learning, signal processing and
communications, with more than 270 SCI-indexed papers and more than 180 IEEE journal papers. Now, he is an editor for several international journal such as IEEE Wireless Communications, IEEE Systems Journal and IEEE Access. His research interests include  machine learning, and low-complexity algorithms with applications in wireless communications.
\end{IEEEbiographynophoto}
\begin{IEEEbiographynophoto}\\
MAOLIN LI is a PhD student in the School of Information and Communication Engineering at Hainan University, Haikou, China. His research interests include massive MIMO, physical layer security, and intelligent reflecting surface.
\end{IEEEbiographynophoto}
\begin{IEEEbiographynophoto}\\
	QINGQING Wu received the B.Eng. and the Ph.D. degrees in Electronic Engineering from South China University of Technology and Shanghai Jiao Tong University (SJTU)
	in 2012 and 2016, respectively. From 2016 to 2020, he was a Research Fellow in the Department of Electrical and Computer Engineering at
	National University of Singapore. He is currently an Associate Professor with Shanghai Jiao Tong University. His current research interest includes
	intelligent reflecting surface (IRS), unmanned aerial vehicle (UAV) communications, and MIMO transceiver design.
\end{IEEEbiographynophoto}
\begin{IEEEbiographynophoto}\\
KE YANG received the B.E. degree from Nanchang University, China, in 2021. He is currently pursuing the M.S. degree with the School of Information and Communication Engineer, Hainan University, China. His research interests include physical layer security and intelligent reflecting surface.
\end{IEEEbiographynophoto}
\begin{IEEEbiographynophoto}\\
BIN DENG received the B.E. degree from East China University of Technology, China, in 2023.  He is currently pursuing the M.S. degree with the School of Information and Communication Engineer, Hainan University, China.  His research interests include massive MIMO, intelligent reflecting surface.
\end{IEEEbiographynophoto}
\begin{IEEEbiographynophoto}\\
XUEHUI WANG received the M.S. degree from Hainan University, China, in 2020, where she is currently pursuing the Ph.D. degree with the School of Information and Communication Engineering. Her research interests include wireless communication, signal processing, and IRS-aided relay systems.
\end{IEEEbiographynophoto}
\begin{IEEEbiographynophoto}\\
FUHUI ZHOU is currently a Full Professor at Nanjing University of Aeronautics and Astronautics. He is also with Key Laboratory of Dynamic Cognitive System of Electromagnetic Spectrum Space, Nanjing University of Aeronautics and Astronautics. He is an IEEE Senior Member. His research interests focus on cognitive radio, cognitive intelligence, knowledge graph, edge computing, and resource allocation. 
\end{IEEEbiographynophoto}
\begin{IEEEbiographynophoto}\\
	CUNHUA PAN (cpan@seu.edu.cn) received Ph.D. degrees from Southeast University, China, in 2015. He is a full professor in Southeast University, China.
\end{IEEEbiographynophoto}
\end{document}